# The Sanction of Authority: Promoting Public Trust in AI

Bran Knowles
b.h.knowles1@lancaster.ac.uk
Lancaster University
Lancaster, UK

John T. Richards
ajtr@us.ibm.com
TJ Watson Research Center, IBM
Yorktown Heights, New York, USA

## ABSTRACT

Trusted AI literature to date has focused on the trust needs of users who knowingly interact with discrete AIs. Conspicuously absent from the literature is a rigorous treatment of public trust in AI. We argue that public distrust of AI originates from the under-development of a regulatory ecosystem that would guarantee the trustworthiness of the AIs that pervade society. Drawing from structuration theory and literature on institutional trust, we offer a model of public trust in AI that differs starkly from models driving Trusted AI efforts. This model provides a theoretical scaffolding for Trusted AI research which underscores the need to develop nothing less than a comprehensive and visibly functioning regulatory ecosystem. We elaborate the pivotal role of externally auditable AI documentation within this model and the work to be done to ensure it is effective, and outline a number of actions that would pro- mote public trust in AI. We discuss how existing efforts to develop AI documentation within organizations—both to inform potential adopters of AI components and support the deliberations of risk and ethics review boards—is necessary but insufficient assurance of the trustworthiness of AI. We argue that being accountable to the public in ways that earn their trust, through elaborating rules for AI and developing resources for enforcing these rules, is what will ultimately make AI trustworthy enough to be woven into the fabric of our society.

## 1 INTRODUCTION

It is commonly stated as fact that trust in AI is important. It is true, more than other digital technologies, AI has a moral valence that demands trust. Fairness, for example, is not merely an abstract concept or guiding principle of "trustworthy AI" [11]; AI has material consequences in the world that affect real people and reflect often longstanding structural inequities. Distrust in this context has "rhetorical power" [20], drawing attention to the ways in which AI fails to live up to our cherished societal values and demanding that we do better.

This is not the reason typically given for the importance of trust in AI. That reason is, in contrast, instrumental: trust is presumed to be necessary for AI adoption (e.g. [4, 18, 21, 26, 30, 41, 46, 48]). Distrust is, therefore, an inconvenience, a barrier to realizing the "promise" of AI through widespread uptake [1, 11]. This neoliberal logic of rational choice in a free market depoliticizes important moral issues surrounding AI and makes it seem as though trustworthiness is secondary to whether consumers can be convinced (even tricked) to adopt it. But this choice is illusory. Can the vast numbers of individuals who distrust or even fear AI [12, 51] really *reject* it? What would that even mean? What member of the public is consciously "adopting" AI? AI is not like a gadget one can pick up off the shelf or an app in an app store. AI is embedded within digital technologies, or even more insidiously, within the digital economy. Very soon all technologies will have some element of AI; every decision will be informed by AI. The general public are not *users* of AI; they are *subject to* AI. Consequently, lack of public trust in AI will not express itself in market forces; public distrust will not function as a corrective toward industry self-regulation of these technologies, impelling companies who develop and utilize AI to do better. So as AI becomes an increasingly unavoidable part of the fabric of society, why should anyone trust it? What guarantee do we have of its trustworthiness?

In this paper, we argue that there is good reason for the public not to trust AI absent the development of regulatory structures that afford genuine accountability. Our focus is on *public trust*, for which understandings of trust appear both underdeveloped and eminently important for the reasons described in the opening to this paper. By "the public" we mean average, relatively non-technical citizens—individuals who encounter AI as an abstruse component of their digital milieu. To date, the field of Trusted AI has focused on developing tools to provide access to indicators of trustworthiness, such as whether outputs are reliable, un-biased, and understandable. A growing body of work seeks to develop resources for a) experts who train AI models and thus require a means of identifying and correcting errors in these models, and for b) decision makers, particularly those in high-stakes decision making contexts, who need to understand the processes that produced the algorithmic outputs they depend on. Such approaches presume a user is forging a trust relationship with a specific, discrete, identifiable AI. This does not describe the relationship members of the public have with AI. As noted above, the public encounters AI often without any aware- ness of having done so, and without any means of assessing its trustworthiness—such is the nature of existence in a world of *pervasive* AI. Even if the members of the public could cultivate greater cognizance of interacting with AI, their ability to form individual trust relationships with any of them diminishes as both the complexity and sheer number of AIs increases. We contend, therefore, that the public can only meaningfully relate to AI as an *abstraction*, what we are calling "AI-as-an-institution", which has important implications for how to succeed in promoting public trust in AI.

This paper will proceed in the following stages. We begin by presenting a brief overview of the dominant theoretical orientation toward trust within Trusted AI literature, and challenge its applicability to public trust in AI. Next, we draw on literature on institutional trust in developing a new framework for promoting public trust in AI. We then describe the pivotal role of externally auditable AI documentation within this model and the work to be done to ensure it is effective, and outline a number of actions that would promote trust in AI-as-an-institution. We end by critically evaluating the role of explanations in promoting trust in AI, and exploring the relationship between AI documentation within organizations and documentation to be developed for external auditing.

By rearticulating institutional trust and the concept of "facework" as it relates to public trust in AI, this paper reveals important distinctions between expert trust and public trust in AI. We seek to correct what we see as a misunderstanding of trust in AI as being equivalent to trust in persons or in individual AI-based systems. This is no mere semantic distinction; it has practical consequences to the promotion of public trust in AI. The institutional nature of public trust, as we show, requires that we look to structural remedies. Specifically, we show that public trust in AI will arise only through development of a robust regulatory ecosystem that provides some guarantee that the public is protected from harmful consequences of AI.

## 2 TRUST IN TRUSTED AI

It is outside the scope of this paper to do a comprehensive treatment of understandings of trust in the Trusted AI literature—this effort is undertaken elsewhere (in preparation). As a general characterization of Trusted AI, however, the field has largely focused on the presumed inability to trust that which one does not understand. Explainable AI (xAI)—within which there are a variety of explanation types [39]—seeks to rectify this information gap by, generally, providing explainees a description of how a model or algorithm inside a "black box" functions [49], or why a particular decision or outcome was reached. This points to a key assumption about trust formation, namely that it has a largely *cognitive* basis [32]. It is this orientation that we turn to now. We focus on a model of trust in AI recently proposed by Toreini et al. [48], which synthesizes and builds on related scholarship, in order to highlight key points of difference that arise in the context of *public* trust in AI. This is not intended as a critique of Toreini et al's model, but rather an urgent cautioning against its applicability to the matter of public trust[1].

Toreini et al. [48] choose the "widely accepted" ABI (Ability, Benevolence, Integrity) framework as a launching point for exploring trust in AI. These three characteristics are thought to be determinative of a trustee's trustworthiness, and the trustor's role is to assess whether a person is capable of delivering against expectations, intends to do so, and adheres to principles in a consistent way that make it likely they will do so [34]. The model proposed by Toreini et al. [48] assumes that ABI (plus Predictability [44]) is generally applicable to the formation of trust in AI, though complicated by an additional layer comprised of *humane qualities*, *environmental qualities*, and *technological qualities* (HET). Humane qualities might include, for example, individual differences in risk tolerance or trusting stance, which make a trustor either more less likely to trust AI, independent of its ostensibly objective trustworthiness. Environmental qualities account for cultural variations between deployment contexts—a sense that "everything is in proper order" ("situational normality") and the existence of various guarantees on trustworthiness ("structural assurances"), such as contracts and regulations, that are conducive to trust [46]. Technological qualities, lastly, "determine the capacity of the technology itself to deliver the outcome as promised" [48]. (Though the authors do not describe it as such, these HET qualities seem to recapitulate ABI—respectively, Benevolence, Integrity and Ability—representing another sense-check on trustworthiness.) An additional temporal dimension is included, recognizing a difference in requirements for initial trust building and maintenance of trust [32]. The authors ultimately use the model to propose that technologies that make contributions to AI fairness, explainability, auditability and safety (FEAS) foster trust in AI by virtue of improving AI trustworthiness—the preeminence of trustworthiness being a near universal feature of Trusted AI literature. Completing the model, FEAS is seen to draw from and implement principles of ethical AI, suggesting that being "ethical" is relevant to determining an AI's overall trustworthiness insofar as it is an aspect of benevolence.

The seminal work of Mayer et al. [34], which originated the ABI model, is now cited over 22,500 times, a testament to its profound influence on understandings of antecedents and outcomes of organizational trust. Trust in organizational contexts is a lubricant of effective cooperation between individuals in working relationships with one another, and while this has been picked up as a means of understanding trust in technology more generally, it's not clear that a model designed to describe trust in that context is readily transferable to a technology context—at least for the purpose of promoting public trust in AI. For one, it is oddly anthropomorphic to confer ability, benevolence and integrity to machines; though this complication is conveniently side-stepped if one assumes that trust in AI is a proxy for trust in the people who developed the AI (an argument questioned by [33]). Further, the model alludes to three stages, articulated [44] as *belief* (formed on the basis of perceptions of ability, benevolence and integrity), *decision* (a willingness to make oneself vulnerable to risk), and *action* (taking on that risk); meanwhile ongoing observation of outcomes informs perceptions of predictability, which circle back to help with calibration of trust beliefs (cf. [43]). Taking these three stages as given, Toreini et al.'s model implies that while an AI's "inherent trustworthiness" [48] may not result in trust for a variety of reasons, trust results when progression through these stages leads one to determine that AI is sufficiently trustworthy for their purposes.

Let us suppose for a moment that this is a largely accurate portrayal of an individual's journey to trusting AI; it is then problematic that members of the public are severely limited in their ability to process trust at multiple points along this path:

---

[1] Note that the authors do not claim their model applies to the public. By neglecting to specify relevant trustors, however, the implication would seem this model should apply in some manner to all trustors.

- *Belief (ABI+):* This presumes a level of expertise on the part of the individual to be able to meaningfully assess trustworthiness and an ongoing commitment to monitoring predictability. As Toreini et al. [48] rightly point out, there are a number of stages in the machine learning pipeline that would need to be considered within such an assessment for one to be in a position to form well-placed trust[2]. The public has neither this expertise nor capacity when it comes to individual AIs, let alone for all the myriad and ever increasing number of AIs in their lives. Moreover, this feedback between observed behavior and belief, which can either strengthen or diminish trust in organizational contexts, is also questionable with respect to AIs given the difficulties the public faces in even recognizing when they are interacting with them.
- *Decision:* The willingness to make oneself vulnerable to risk presumes that one is sufficiently capable of understanding what those risks are. Moreover, "risk" is too simplistic for this context. "Deciding" whether AI is generally good for society is qualitatively different from, for example, deciding whether one opens oneself up to too great a risk in relying on a colleague to perform a specific task. While Toreini et al's [48] model clearly seeks to account for ethical deliberations, it is not clear that such deliberations lend themselves to decisions premised in notions of risk.
- *Action:* Mayer et al's [34] model clearly implies a choice to act or not act. If by "act" one means adopt or consent to be subject to AI, the public cannot be said to make such choices. Toreini et al. [48] attempt to resolve this by noting that there can be dissonance between beliefs, decisions and actions, resulting in "*as if* trust" [54], though this leaves unresolved the issue of why, if the final result is the same, trust beliefs or decisions even matter. Also, to the extent that accountability relies on parties being able to act (e.g. sanction untrustworthy actions on the part of the trustee), the irrelevance of action here removes a critical "mechanism to facilitate better behavior" [53] (see also [7, 14]), which would otherwise feed a virtuous cycle for trust.

For these reasons, a new model is needed to account for public trust/distrust in AI. In what follows, we provide a foundation upon which to construct a more appropriate understanding for this altogether different context of pervasive AI, where the assessment of trustworthiness simply does not scale.

## 3 TRUST IN INSTITUTIONS

In broad terms, our concern is to do with the issues of scaling human-computer trust to our technologically "complex" lives. This problem holds for ubiquitous computing generally, and it is surprising that models of trust in ubiquitous computing have focused more on issues to do with people being separated in space [13, 28, 29, 45] than accounting for aspects of "systemness". We focus here on how individuals' trust (or distrust) of AI is based in its "systemness", thereby overcoming the problem of not being able to forge trust in specific AIs.

We need not invent novel trust mechanisms here. Sociological scholarship provides useful insights into how humans readily manage trust in such circumstances, having already grappled with the problem of trust in late modern societies. Because the effort required to forge and manage interpersonal trust relationships does not scale to the level of social complexity in these societies, a different basis for social order emerges from "system trust (i.e., trust in the functioning of bureaucratic sanctions and safeguards, especially the legal system)" [31]. Individuals do not develop trust in these systems through careful and ongoing assessment of their trustworthiness; instead, *one trusts that the system itself has appropriate mechanisms for ensuring trustworthiness* [17, 27].

Our discussion is rooted in Giddens' structuration theory[3], which understands societal structures (e.g. expert systems, or institutions) as "both medium and outcome of the practices they recursively organize" [16]. In other words, "systemness," as we described it above, is something that is dynamically negotiated through dimensions of *structure*. "Structuration", i.e. this interplay between structure and its reproduction, is often used synonymously with "institutionalization" [5], and we find it useful to speak of trust in "AI-as-an-institution" as a means of signalling a key difference between the kind of trust that is relevant to the context of pervasive AI (as the public experiences it) and that which experts might have in individual AIs. But to be perfectly clear, our use of the phrase "institutional trust" should not be confused with trust in the organizations that produce or employ AI. The institution is the wider ecosystem that determines the rules that these organizations abide by and the resources for enforcing those rules; so trust in AI-as-an-institution would be based in a perception that *the system creates conditions for organizations to develop sufficiently trustworthy AIs and to employ them responsibly.* This means, as we will expand on later in the paper, that the inverse is also true: if the AIs that organizations develop and deploy are seen to be untrustworthy, there has been an institutional failure to effectively regulate AI.

Another way of explaining this is to say that when one is trusting an abstract system (such as an institution), one is trusting in the system's *structures*, i.e. "the rules and resources that govern its working and its continuous reproduction in the form of regular social practices" [27]. So what are these dimensions of structure that comprise institutions? Giddens refers to these as *signification, legitimation, and domination* [17]:

> **Signification** refers to the implicit language of symbols and symbolic actions that derive their meaning in reference to the institution. A classic example is a white coat that signifies a doctor [25]. People draw on their understandings of how the world works (e.g. doctors wear white coats; doctors exist in hospitals; a person in a hospital wearing a white coat is a doctor) to infer meaning from these symbols.
>
> **Legitimation** provides norms (i.e. rules that govern appropriate actions) and values which are known to agents in the institution. These norms and values are made visible and are reinforced by institutional sanctions of failures of its members to embody these expectations [27].

---

[2] Toreini et al. [48] acknowledge limitations of individuals' capabilities with respect to assessing ability and benevolence, and propose that individuals accomplish this indirectly through assessment of the ability and benevolence of the entity developing the AI. This is similar to, but does not go as far as, the model we propose in this paper.

[3] See [25] for an overview of uses of structuration theory in information systems research.

**Domination** refers to the ways that power can be applied, which depends on the degree to which agents within the institution are empowered. Agents are empowered through a) allocative resources, which grant them certain control over material objects, and through b) authoritative resources, which grant them power over other agents [27].

Trust arises in part from a perception of coherence between these dimensions [27]—resonating with scholarship on trust that emphasises the importance of *integrity* [34]. More than this, however, each structure is subject to breakdown so must be seen to be functioning in ways that ensure trustworthiness of the system. If individuals breaking the rules are not punished for doing so, the norms and values claimed by the institution lose their legitimacy. If symbolic actions are subject to wide ranging interpretations, these lose their communicative value. Finally, agents need to be seen as sufficiently empowered in their roles to be able to exert power; and at the same time, trust is stronger when power is seen to be subject to checks and balances provided by the system itself, rather than relying on the benevolence of powerful individual agents [27].

The question remains as to how trust in institutions grows. For this, we turn to the concept of *facework*, originally developed by Goffman [19] in relation to self-presentation, reformulated by Giddens [17] with respect to concerns about trust in abstract systems, and rearticulated by Kroeger [27] to explain the development of institutional trust. Goffman defines *face* as: "the positive social value a person effectively claims for himself by the line others assume he has taken" [19]; thus *facework* is "actions taken by a person to make whatever he is doing consistent with face" [19]. Giddens conceives of facework as a mechanism for building up trust in "faceless commitments" (i.e. "abstract systems") [17]—its function is to bridge the gulf between the mechanics of trust operating commonly between individuals, which are "sustained by or expressed in social connections established in circumstances of copresence" [17], and trust that operates in socially "disembedded" contexts. This is accomplished, Giddens argues, "by experts and other representatives at the *access points* of these abstract systems" [17]. Building on Giddens, Kroeger rearticulates facework as "the translation of interpersonal trust into trust which pertains to an institutional system, based on the conduct of system representatives who, using their agency, are seen to draw on the system's rules and resources in devising behaviours able to signal trustworthiness" [27] (italics removed). In order to promote trust, therefore, representatives of the institution need to "visibly draw on, and reproduce, dimensions of structure which determine the trustworthiness of the institutional system" [27]. Or in other words, those who are seen to be part of the institution foster trust by demonstrably upholding the values and norms of that institution, particularly with reference to the power that oversees their normative behavior.

## 4 DEVELOPING A MODEL

Drawing from this structurationist understanding of the basis for trust in institutions, we now endeavor to provide a model of public trust in AI-as-an-institution. It should be noted at the outset that we do not contend that AI-as-an-institution has been deliberately designed to have particular structural qualities—indeed, we suggest that a failure to consider the institutional nature of public trust in AI has left these dimensions severely underdeveloped. This, however, does not undermine the legitimacy of the model. While signals of institutional trustworthiness can be intentionally "given" through purposeful representation, they are also "unintentionally 'given off'" [27], and observers cannot help but infer structural qualities of the institution [27]. It is our thesis that lack of public trust in AI has little to do with people's inability to understand how AIs work; rather it is a response to an awareness of a lack of structural assurances of the trustworthiness of the AIs pervading society.

### 4.1 Structures

In understanding the public's tendency to distrust AI, it is important to look at what the public believes AI is and does and how these beliefs affect attitudes toward these technologies; how this relates to what AI actually is and does; and how a mismatch between the two may have arisen. The first thing to note of AI is that whatever it is now, it has been built up in the public imagination as a thing to be feared, a cautionary tale of human hubris. Perceptions of AI are "informed by long-standing cultural imaginaries of machines that escape the control of their creators, and the promises and perils of automata and artificial life" [12]. First impressions, which are considered so important to trust formation [32, 35, 36], have already been made, and they are not positive. This means that the public begins from a trust deficit that must be overcome, perhaps by means not required of other contexts of technological trust building (cf. [32]). Fears of AI are exacerbated, Elish & Boyd argue [12], by hyping (in order to sell) the technology. It is frequently marketed as working "like magic", feeding a sense that AI is beyond human comprehension and thus unaccountable to its creators. Anthropomorphic language also plays a part in fueling anxieties. Meant to translate technically complex functions into comprehensible behaviors, the language used in marketing and media imbues AI with human capabilities it does not in fact have, such as "thinking"—the effect of which is to present AI as having agency beyond the control of its human developers, playing directly into the science fiction narrative of AI's inevitable domination of the human race. Media coverage of scandals such as Cambridge Analytica portrays AI as a weapon wielded by the powerful to hijack democracy through AI-enabled mind control. This narrative has attached itself to AI: "As with any potent frame, the story that an AI system has been built to manipulate the public's hive mind is hard rhetoric to combat" [12]. And public spectacles such as the Facebook Congressional hearings make it clear that elected public officials are so limited in their technical knowledge that they can hardly be trusted to craft meaningful legislation in this space nor mete out sanctions against companies found to be acting unethically.

It is against this backdrop that structural questions gain salience. **Legitimation:** Who or what is determining the values that might guide the development of AI toward moral ends? What visible sanctions have been applied that might reveal a normative culture within the AI sector? Are my own values reflected in that culture? **Domination:** What laws are in place to protect the public from harm? Are they enforceable? Who or what is empowered to sanction AI? To whom are AIs ultimately accountable? **Signification:** What would be indicative of ethical or trustworthy AI? How do we know when we see unethical AI? How do certain features of AI translate to social harms?

These questions, of course, are not new. What the structurationist lens provides is a new way of understanding the importance of ongoing work in the area of Trusted AI. **Legitimation:** The avalanche of corporate, non-governmental and (inter-)governmental statements on principles to guide ethical AI [24, 38, 47] are attempts at articulating *face* by defining a set of positive values claimed by AI-as-an-institution. Where they fall short is in prescribing or even indicating normative implementation of these principles—a shortcoming much lamented in the literature [52]. **Signification:** The exigence of AI documentation methodologies and toolkits are attempts at rendering AI features in a way that visibly reflect an AI's adherence to trustworthy AI principles, thereby actively constructing a language for representing abstract social constructs such as "fairness". **Domination:** Calls to develop a regulatory ecosystem for trustworthy AI [11] indicate a growing awareness of the need for external auditing of AI by technically competent regulators who are empowered to enforce laws and regulations. The work of developing this ecosystem remains to be done.

In what follows we focus on AI documentation as a key to enabling this regulatory ecosystem, thus providing the necessary conditions for public trust in AI to grow.

## 4.2 AI documentation to date

Documentation has long been essential practice within software engineering, and it is no surprise that AI would need to be thoroughly documented. Recent efforts to standardize AI documentation capture information related to the safety and performance testing an AI has undergone [3, 22], performance characteristics and intended use cases of algorithms [37, 42] and details of algorithms' training datasets [6, 15, 23] that are relevant to determining the contexts within which the AI can be responsibly applied. These differ from standard software engineering documentation in that they are designed "to address ethical and legal concerns and general social impacts of such systems" [42] by providing transparency on key facts regarding the AI's development.

AI documentation is ideally (if still rarely) used within internal governance practices across the whole AI lifecycle. As an artifact, documentation can improve communication within a team—in particular regarding team members' individual impacts on the resulting product—and can be used to illustrate and argue the need for additional resources to ensure the AI lives up to ethical standards; and as a process, it encourages individuals to consider ethical implications (either directly or indirectly through associated technical evidence) throughout the product's development [40]. In short, AI documentation helps ensure that the AI being developed conforms to *organizations' own values and standards*.

While there are notable exceptions for the use of AI in already highly regulated sectors (e.g. finance), it is generally true that AI documentation has not been developed for the purposes of external auditing. No doubt AI documentation is informed by emerging guidelines for trustworthy AI, but compliance with these guidelines is entirely voluntary. Certainly companies have an inherent drive to avoid reputational damage and might seek to self-certify their trustworthiness through processes of documentation[4], but for reasons outlined in the introduction, this is not a sufficient basis for trust in AI-as-an-institution. Public trust in AI will emerge when power resides in the institution itself, rather than in the power (and benevolence) of a number of organizations committed to trustworthy AI. And having now described the structural conditions which support trust in institutions, AI documentation in its current form does not provide sufficient structural guarantees of AI's trustworthiness. In what follows, however, we argue that AI documentation can play a critical role in impelling compliance with sector-wide guidelines and legislation, and can indeed be the appropriate mechanism for AI-as-an-institution to ensure trustworthiness of individual AIs.

## 4.3 The role of algorithmic documentation in promoting public trust in AI

As critical context for our proposed approach to promoting trust in AI-as-an-institution, we must first note certain unique challenges in both documenting and regulating AI (which are, of course, intertwined). In principle, AI products and services are subject to existing legislation including consumer protection, product safety and liability, non-discrimination, and data protection (e.g. GDPR) [11]. These laws relate to the majority of public concerns around AI, and might provide decent institutional guarantees on trustworthiness were it not for the fact that these laws are notoriously difficult to enforce when it comes to AI. Certain characteristics of AI ("opacity ('black box-effect'), complexity, unpredictability and partially autonomous behaviour" [11]) present major challenges in evidencing legal compliance and non-compliance. In terms of non-discrimination law in particular, "AI can use unintuitive proxy data to make decisions about humans that do not make our 'alarm bells' ring" [9]. The nature of AI being "abstract and unintuitive, subtle, intangible, and difficult to detect" makes it less likely people will seek legal recourse for discrimination [50], thus undermining the legal safeguards ostensibly afforded by the system. Many of these same issues pertain to so-called "soft-law" [24] (i.e. various ethical/trustworthy AI guidelines): what does implementation of principles such as justice, fairness, non-maleficence, and privacy [24] look like, and how will indicators of their successful implementation be detected?

A major implication of understanding AI as an institution, implicit in this paper up until now, is the need to develop a means of systematically weeding out AIs deemed harmful (or risky) so that what remains is trustworthy. Because existing laws have limitations in this regard, the need for AI-specific legislation that can be more "effectively enforced" has been raised [11]. Surely, in order that such legislation *has teeth*, as the saying goes, one first needs something to *chew on*. And this is where AI documentation comes in: documentation artifacts could provide external auditors with facts needed for determining the trustworthiness of a given AI in order to exercise authority (**domination**). AI-as-an-institution cannot visibly draw on rules that do not yet formally exist. Thus, agreeing on what might constitute a "fact" is also critical (**signification**); as would be the establishment of a set of normative decisions about what constitutes legal and ethical non-compliance (*legitimation*). Systemic trustworthiness would result from integrity across these interlocking dimensions of structure. Figure 1 provides an overview of this basic model for the promotion of trust in AI-as-an-institution mapped to these structures.

---
[4]Note that efforts such as Denmark's Data Ethics Seal [10] and Malta's voluntary certification system for AI [2] are premised in this same notion of the reputational advantage of trustworthiness.

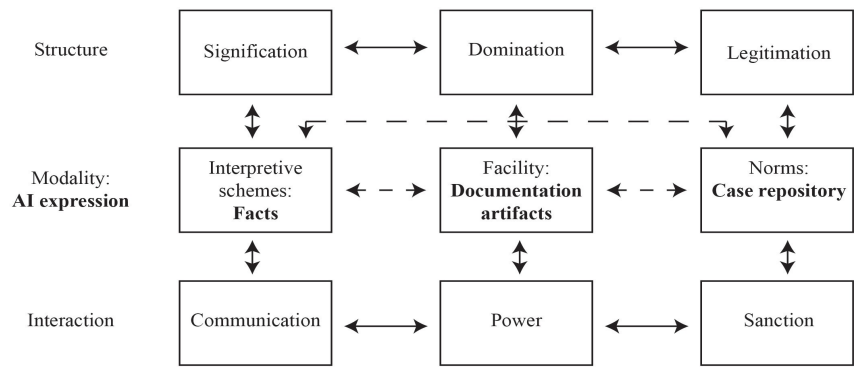

**Figure 1: Dimensions of structure and their expression in AI context. Adapted from Jones & Karsten, 2000 [25].**

|  | **Positive Facework** | **Negative Facework** |
| --- | --- | --- |
| **Domination** | Emphasizing control over AI; referencing documentation (and external auditing of it) | Promisory rhetorics; public-facing explanations |
| **Legitimation** | Celebrating good AI and the elimination bad AI; displaying certification (referencing rigor of certification process) | Brushing untrustworthy AI under the carpet; multiplicity of self-certifications by orgnizations |
| **Signification** | Explaining what the institution means by "trustworthy AI"; relating this definition to public concerns | Emphasizing purely technical reasons AI is trustworthy; referencing inconsistencies in interpretations of fairness |

**Table 1: Examples of positive and negative facework for AI-as-an-institution. Adapted from Kroeger, 2017 [27].**

This model offers a new way of envisaging the potential of AI documentation to promote public trust in AI. While sometimes likened to food nutrition labels [23, 40, 42], there is little reason to believe that members of the public would be able to interpret documentation to make informed decisions about using or being subject to AI. But members of the public do not need to trust individual AIs at all (cf. [8]); what they need instead is the sanction of authority provided by suitably expert auditors that AI *can be trusted*. In order for AI documentation to be interpretable by these auditors, a shared language would need to be established for relating evidence (facts) to underlying legal and ethical constructs. This likely includes the development of both "a standard set of statistical evidence" [50] for criteria such as fairness—work that is making notable progress in mapping various interpretations of fairness to computed metrics —as well as norms for annotating documents with relevant contextual information (cf. "contextual equality" [50])[5]. Previous work has proposed a methodology for producing AI FactSheets [42] that convey necessary information to an intended audience, and an obvious next step would be working with individuals who are sufficiently knowledgeable of the legal and ethical requirements to understand what actionable insights they can glean from FactSheets, perhaps even co-designing FactSheets that are specifically tailored to the needs of future regulators. Once standards are established at the sectoral level for capturing and representing facts, a next step would include initiatives that seek to enhance skills of "sectoral regulators...to effectively and efficiently implement relevant rules" [11].

We propose a further need to establish something akin to case law—i.e. decisions regarding the trustworthiness of specific AIs— that can be referenced within the documentation and regulation. There is need of active deliberation in the AI community regarding certain AIs that could serve as precedents for both "trustworthy" and "untrustworthy" AI[6]. This is essential in establishing institutional norms, but would also help regulators make satisfactorily consistent determinations about the trustworthiness of a given AI, thereby *reproducing* these norms. Documentation could, for example, point regulators to related exemplars, possibly annotated with details of how they differ from these established cases.

---

[5] This may build on the efforts of creating documentation that communicates intended use cases [6, 15, 37, 42].
[6] As an initial step—i.e. to help catalyze progress toward being able to regulate and thus make decisions on real AIs—leadership within the AI community could develop a series of hypothetical AIs as case studies.

We imagine these components as needing to co-evolve (see Figure 3). Normative decision making would be built into the documentation artifacts, which would begin to develop effective ways of referencing precedent; the language for communicating facts about an AI would evolve as new exemplars highlight the need to capture additional or different facts; the documentation itself, by virtue of what can actually be captured, will shape the way that facts are communicated; and what facts can be captured will determine what technical features within AIs prompt sanctions.

One final element of our model remains, namely what positive facework (i.e. behaviors of representatives of AI-as-an-institution that promote public trust in AI) might look like. While this is not an exhaustive list, the following are indicative behaviors to en- courage those who are seen to represent AI-as-an-institution (This includes, possibly most conspicuously, the organizations developing and deploying AI, but also individuals working at these organiza- tions, legislators and inter-governmental bodies who determine the rules, and regulatory bodies who enforce these rules). **Domina- tion:** A necessary first step would be contributing to a sense that AI is within human control, both to help the public appropriately calibrate their fears and begin to dismantle longstanding cultural imaginaries. To this end, positive facework would include intention- ally eschewing "promissory rhetorics" of what AI might one day do [12] for greater clarity regarding current capabilities and inherent limitations, communicating how its "intelligence" is constrained to incredibly specific contexts for which its performance has been optimized. Positive facework could even be as simple as making it clear that AIs are being documented. The fact that AI is being documented means that someone with appropriate skills could examine the AI for trustworthiness. **Legitimation:** Other positive facework would include making an effort to publicly acknowledge when harmful AI is detected and discontinued. This would serve to reinforce a sense that the system is controlling for trustworthiness, while also helping articulate the norms that constitute trustworthy AI. Counterbalancing this with stories of AI doing good in society would prevent against perceptions that AI is threatening by default. AI certification could be a form of positive facework if it represents a genuine auditing process. Organizations whose AI receives certification can do facework by articulating their own commitment to ethical and trustworthy AI and proudly displaying credentials, but even more effectively, by emphasizing the credibility of the underlying certification process. **Signification:** There is a need, more fundamentally, to explicate how concerns underlying public distrust of AI are accounted for within the regulatory environment. This would mean clarifying what legislation and/or regulators are controlling for, what kind of indicators auditors look to as evidence in determining trustworthiness (including guarantees of fact authenticity), and overall why the resulting AI is "trustworthy". In this paper we have conspicuously avoided defining "trustworthy", in part because this is something the community must feel its way toward through co-evolution of these dimensions of structure (see Figure 2 showing the relationship between trustworthiness and public trust in AI). As this understanding of trustworthy AI evolves, not only does a shared institutional definition of "trustworthiness" need to be articulated, but it needs to be articulated in reference to public concerns about AI. In other words, important facework would be communicating how the system ensures that AI is trustworthy *in a way that meaningfully relates to how the public would define trustworthy AI.*

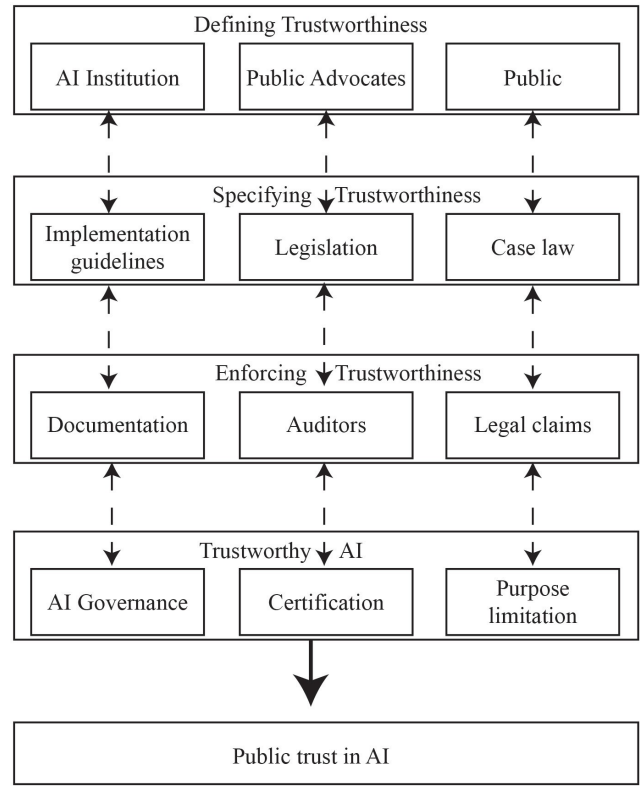

**Figure 2: Stages in promoting public trust in AI.**

## 5 DISCUSSION
### 5.1 Reconsidering explanation

There is a common, but generally unspoken, assumption that explanations of an AI's inner workings are essential for promoting trust. One of the key insights emerging from our model is that xAI actually has no direct role in promoting *public* trust in AI. Certainly, if one cannot explain *to relevant expert stakeholders,* how an AI is arriving at a decision, that AI should not be allowed to be part of the decision making process. We see this playing out, for example, in heavily regulated industries such as finance. The ability to regulate depends, therefore, on being able to document key aspects of an AI in ways that render it interpretable to an auditor—the xAI toolset and fundamental insights generated from the body of xAI may be applicable to this challenge. But critically, while it is important that somewhere an element of structure accounts for what is happening inside the black box (e.g. in documentation artifacts), public-facing explanations of inner workings would not be an example of positive facework; indeed, we would go so far as to propose such explanations are negative facework and would

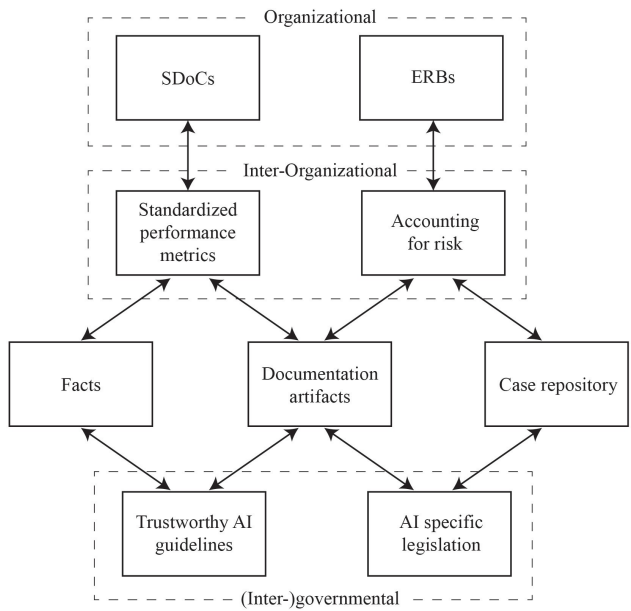

Figure 3: Co-evolution of structural dimensions and the role of organizational documentation practices (SDoCs and ERBs).

## 5.2 Beyond component level documentation

It is worth stating explicitly that AI does not provide total solutions; rather it provides intelligent *components* of systems. Concern over these intelligent components often obscures the overall systems in which these AIs operate. What matters to the user of any system is not *how* the system works, but *that* it works. Users do not worry about the chips in their computers because it is irrelevant to their operating of the machine. Similarly, AI documentation would need to provide a fundamental assurance that the AI meets standards sufficiently well to be used within a system. Comparing AI documentation with a Supplier's Declaration of Conformity (SDoC) makes sense for this limited purpose [3, 22]. As with semiconductor standards that ensure that chip-level components will operate within certain environmental limits and provide reliable functionality as specified, documentation would need to show how the AI meets the requirements of those integrating AI components within their systems.

But this raises an important question: If an AI component contributes to a decision, should it matter that the AI component is biased if that bias is compensated for by a human in the loop? The AI bias would absolutely matter to the person whose job it is to integrate that component into a system, as they would need to devise ways of handling that bias[7]. But for the recipient of a decision, such as a member of the public, the system could be trustworthy even while the AI lacks guarantees of trustworthiness, *and vice versa*—a perfectly functional component does not guarantee a trustworthy system. By "communicat[ing] performance levels of the service…in a standardized and transparent manner" [3], SDoC's make it more likely that AI systems will be developed to be trustworthy because organizations developing AI systems would have the necessary information to purchase trustworthy components and integrate them appropriately. As such, they would be an integral part of a trustworthy AI ecosystem; but do 'chip level' matters (e.g. accuracy with respect to a test dataset) meaningfully relate to public concerns about AI? The public, after all, is not worrying about the components; they are worrying about the systems that use AI.

Ethics review boards are beginning to emerge within large organizations, and have a different set of concerns than those that a system integrator might have. Typically, these boards are driven by concerns about *risk*, reporting to an organization's Chief Risk Officer. What an ethics review board would need to see in terms of documentation is both different to a SDoC and closer to the level of lay public concerns about AI. (When it comes to documentation, one size does not fit all.) While again these ERBs contribute to the trustworthiness of AI, we are not suggesting that ethics review boards provide sufficient structural assurance for public trust in AI, as these processes are still internal to a given organization—though in developing documentation artifacts that could be used for the purposes of external auditing by independent regulatory bodies, they might serve as a good starting point for considering AI impacts beyond the component level. As critical as it is to develop documentation that supports external auditing, however, documentation practices internal to organizations developing AI make important erode trust in the system. Attempting to explain to the public how an AI works might subtly imply that it is necessary for individuals to understand how they work—either so that the public can make their own decisions about which AI is trustworthy, or so that they can "challenge the operation and outcome of a model" [48]. These are functions that ought to be fulfilled by the institution itself to ensure trustworthiness so that members of the public do not need any technical knowledge of AI to be able to live their lives both unconcerned and unharmed by it.

We are not, however, claiming that the public should not be able to scrutinize or challenge an AI. In fact, lawsuits brought by the public or public advocates would be an important force for advancing norms (i.e. helping to grow a 'case repository') for AI. But as noted by Wieringa [53], "The transparent workings of a system do not tell you why it was deemed 'good enough' at decision making, or why it was deemed desirable to begin with…Nor does it tell us anything about its specifications or functions, nor who decided on these, nor why." This is the sort of information that would be provided by documentation artifacts, which would more appropriately service public accountability than a necessarily simplistic model of how the AI works (cf. [39]). Another way of saying this is that explanation is insufficient for accountability; the public does not need to know how an AI works in order to trust it, but rather needs to know that someone with the necessary skillset for understanding decisions made throughout the AI lifecycle has the resources needed for examining AI and the authority to enforce decisions regarding its trustworthiness.

---

[7] A simple example of compensating for untrustworthy AI components would be to develop a workflow that specifies to ignore a prediction if confidence drops to a certain level.

contributions to promoting AI that is worthy of public trust (see Figure 3).

# 6 CONCLUSION

This paper provides an overview of a theoretical framework that accounts for the distinct institutional nature of public trust in AI and the new role documentation could play in fostering public trust. There are many facets of this argument that could be extended in future work, and it is our sincere hope that others join us in developing this further. We lay out the high-level argument in full here to illustrate the need to reorient Trusted AI research. The prevalent misunderstanding of trust in AI as being equivalent to trust in persons or in individual AI-based systems promotes a potentially irresponsible notion: that individual user-consumers are capable of acting as the arbiters of which AIs are sufficiently 'trustworthy'. It is not the public's responsibility to keep AI honest. This should be the responsibility of specially trained individuals who are skilled enough to make determinations about trustworthiness—not only because of how difficult it is to assess the risks and harms of AI, but because potential harms are implicated at the societal rather than individual level.

This means that fostering public trust in AI is not simply a case of mastering Explainable AI; nor it is simply a case of standardizing various performance metrics for AI components and developing internal AI governance practices. Organizations developing AI have a responsibility to implement AI components appropriately and ensure their systems live up to ethical standards, but public trust requires some authority that impels organizations to take these responsibilities seriously and to validate their interpretations of these standards. Public trust requires the development of a wider infrastructure that creates conditions for *worthy* AI to thrive: an entire system of rules needs developing, alongside resources that empower suitably-skilled individuals to enforce those rules. Succeeding in this undertaking will take intensive interdisciplinary collaboration in translating still under specified principles of trustworthy AI into standards that can be satisfactorily accounted for within all systems that include AI components.

AI documentation is clearly implicated in this strategy as a resource to be used by auditors in assessing trustworthiness (or more officially, compliance); but this is to radically reconceive of the value of documentation. There is growing interest in how AI documentation can capture standardized performance metrics, but if underpinned by notions of trust as being formed between a user and an individual AI-system, there is a risk that documentation will be used merely to whitewash AI, for example as part of a new 'terms and conditions' style consent mechanism. Our framework provides a clear focus and exciting impetus for research on algorithmic documentation, namely that it can serve as the lynchpin of a regulatory ecosystem that accounts for and ensures AI trustworthiness.

We have argued in this paper for the need to undertake[8] a major effort towards developing standardized documentation that empowers external auditors to sanction AI in accordance with public notions of trustworthiness. As challenging as this will no doubt be, there is no shortcut to public trust in AI. It is tempting to focus on the component level, to work on making sure that AI is trustworthy whether or not the public believes it so. (After all, the public has no choice but to 'accept' AI, even if they distrust it.) But this is to miss a key point about the relationship between trust and trustworthiness: seeking to earn trust forces one to become more trustworthy. Ultimately, being accountable to the public in ways that earn their trust is what will make AI trustworthy.

---

[8] This does not conflict with the argument in [42] that documentation standards will be somewhat different for different stakeholders in different organizations and different industry sectors, but does suggest that there is an inherent tension between standardization and customization that needs to be brought into balance over time.

## ACKNOWLEDGMENTS

This work is partially funded by the ESRC funded grant BIAS: Responsible AI for Labour Market Equality (ES/T012382/1) and by the Data Science Institute at Lancaster University. The authors would like to thank Richard Harper for his enormously helpful comments on an early draft of this work.